# The Edwards Accelerator Laboratory at Ohio University

Zach Meisel*, C. R. Brune, S. M. Grimes, D. C. Ingram, T. N. Massey, A. V. Voinov

*Institute of Nuclear and Particle Physics & Department of Physics and Astronomy, Ohio University, Athens, Ohio 45701, USA*

**Abstract**

The Edwards Accelerator Laboratory at Ohio University is the hub for a vibrant program in low energy nuclear physics. Research performed with the lab's 4.5MV tandem accelerator spans a variety of topics, including nuclear astrophysics, nuclear structure, nuclear energy, homeland security, and materials science. The Edwards Lab hosts a variety of capabilities, including unique features such as the beam swinger with neutron time-of-flight tunnel and the integrated condensed matter physics facility, enabling experiments to be performed with low-to-medium mass stable ion beams using charged-particle, gamma, and neutron spectroscopy. This article provides an overview of the current and near-future research program in low energy nuclear physics at Ohio University, including a brief discussion of the present and planned technical capabilities.

*Keywords:* Type your keywords here, separated by semicolons ;

## 1. Lab Overview

The Edwards Accelerator Laboratory (EAL) primarily consists of a 4.5MV T-type tandem Pelletron which feeds six target stations at present. Several of the target stations are equipped with dedicated set-ups that are suited to particular experimental techniques, including neutron spectroscopy, charged-particle spectroscopy, and material characterization, whereas others serve as multi-purpose target stations in order to maintain operational flexibility. A schematic of the EAL is shown in Figure 1.

* Corresponding author.
  E-mail address: meisel@ohio.edu

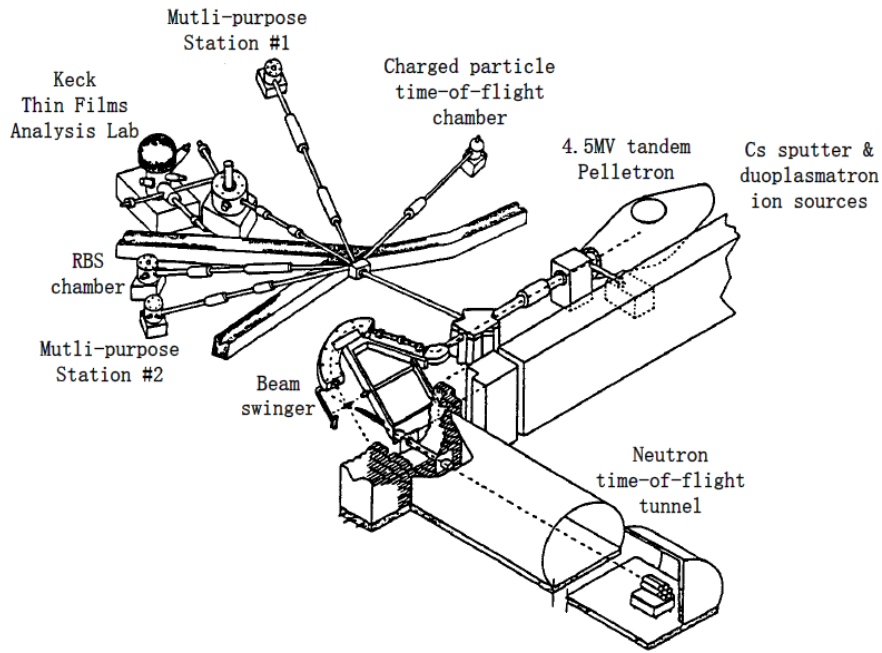

Fig. 1: Schematic of the Edwards Accelerator Laboratory. (Modified from Howard et al. (1981).)

Over the 46 years of operation of the EAL tandem, experiments performed at the lab have led to significant progress in a wide variety of topics, including spectroscopy via neutron elastic scattering, reaction mechanism investigations using charged-particle evaporation, level-density determinations employing a variety of techniques, and material characterization for condensed matter physics studies. The research profile of the EAL contains a particular emphasis on employing neutron beams and neutron detection, leveraging the lab's unique capabilities, such as the neutron time-of-flight tunnel (See Section 3). Another notable feature of the EAL is the Keck Thin Films Analysis Facility (See Section 5), which enables surfaces and thin-films to be prepared and then analyzed using the multitude of material characterization tools provided by ion beams, all while under ultra-high vacuum conditions

The following sections detail major components of the EAL. Section 2 discusses beam production and acceleration, Sections 3-6 discuss existing dedicated target stations, and Section 7 discusses multipurpose target stations and plans for upgrades to the EAL anticipated in the near future.

## 2. Beam Production and Acceleration

Ion beams are produced with a single-cathode cesium-sputter ion source or, for $^3$He and $^4$He, with a duoplasmatron ion source. Typical ion beams include stable isotopes of hydrogen, helium, lithium, and carbon (though heavier ion beams are possible) with ~µA direct-current beam intensities. For experiments requiring pulsed beams, beam-bunching is available with ~ns timing resolution.

Ion acceleration is provided by the terminal of the T-type tandem accelerator. Following the upgrade to a Pelletron charging system five years ago, experiments have successfully been performed using terminal voltages ranging from ~1MV up to 4.6MV. Ions are charge-stripped by a thin carbon foil, though higher charge states can be achieved by employing the gaseous stripper installed in the accelerator terminal. Ions are then directed to one of six target stations using various dipole switching magnets.

## 3. Beam Swinger and Neutron Time-of-flight Tunnel

The beam swinger and neutron time-of-flight tunnel (Finlay et al. (1982)), shown in Figure 2, enable the production of quasi-monoenergetic neutron beams (~500keV to ~25MeV, Brune et al. (2011)) for material characterization and neutron detector commissioning (Massey et al. (1998), Hall (2001), Peters et al. (2016)), as well as high-precision (~few keV) spectroscopy and reaction-mechanism explorations using neutron-emitting reactions. The combined swinger and tunnel obviate the need for a large neutron-detection array, instead allowing the beam to be rotated and outgoing neutrons to be detected with a well characterized, fixed detector system.

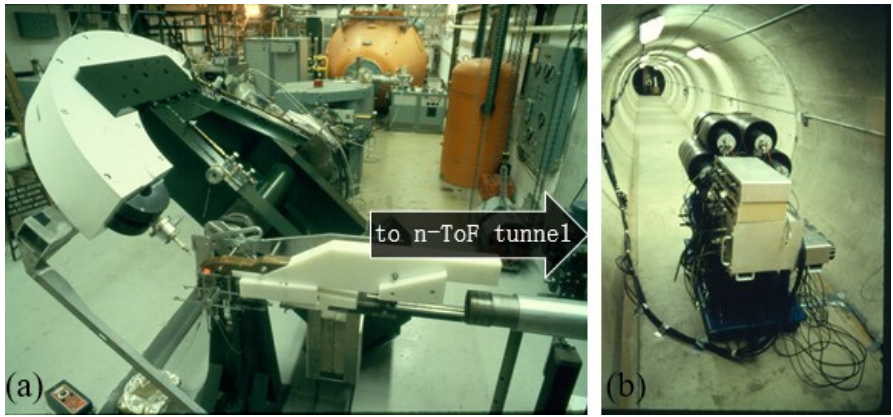

Fig. 2: (a) Beam swinger and (b) neutron time-of-flight (n-ToF) tunnel.

The beam swinger, which originated at Michigan State University (Bhowmik et al. (1977)), is capable of providing an incoming beam angle between -4°≤$\theta_{lab}$≲180° and a beam spot size at the target location on the few-millimeter scale. Both solid and gaseous targets can be used. Notably, tritium targets are available both for 14+ MeV neutron beams (from D+T) and for studying tritium-burning reactions that are of relevance for Big Bang nucleosynthesis and internal confinement fusion diagnostics (Parker et al. (2016)).

The neutron time-of-flight tunnel is a ~30m-long, ~2m-diameter concrete tunnel buried under ~1m of Earth. During experiments, the only opening to the tunnel is a modular collimation system along the beam axis. The amount of collimation can be controlled by using different combinations of cylinder-like polyethylene components. Neutron detection, which is typically performed by fission chambers, an array of NE-213 scintillators, or lithium glass detectors, can be performed anywhere from 4 to 30m from the target location.

The beam swinger and neutron time-of-flight tunnel have been used for a wide variety of experiments. Such measurements have contributed to developments in applied nuclear physics, including internal confinement fusion (Parker et al. (2016)), medical physics (Howard et al. (1996)), and homeland security (Cooper et al. (2013)), nuclear reaction mechanism studies (Mishra et al. (1994), Ramirez et al. (2013)), investigations of nuclear structure for astrophysics (Parpottas et al. (2004), Parpottas et al. (2005)), as well as neutron scattering cross section determinations (Resler et al. (1989)).

## 4. Charged-particle Time-of-flight Spectrometer

The charged-particle time-of-flight spectrometer, shown in Figure 3, is a cylindrical target chamber with ports for evacuated arms to be attached at various angles, each capable of having a silicon surface barrier detector mounted at its end (Wheeler (2002)). Typical arm lengths are between one and two meters, depending on the required time-of-flight resolution (where the present detector timing resolution is ~90ns) and outgoing particle energy for a given angle. The arm ports are mounted at angles ranging from 22.5° to 157.5°, where the measured angles for a particular experiment depend on the details of the reaction being investigated (Voinov et al. (2007), Ramirez et al. (2015)).

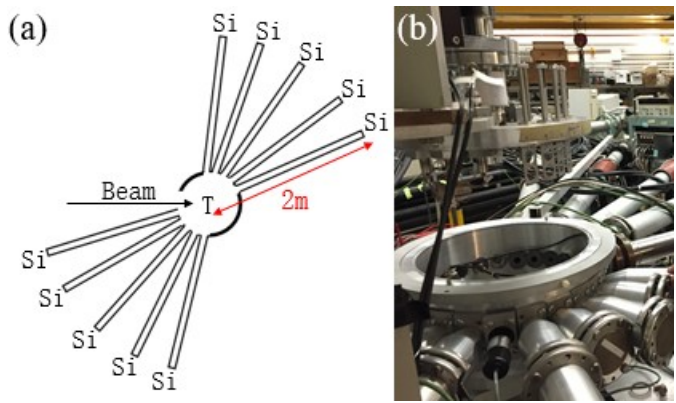

Fig. 3: Charged-particle time-of-flight station (a) schematic and (b) image in the configuration of Voinov et al. (2007).

The charged-particle time-of-flight spectrometer has primarily been used to measure charged-particle evaporation spectra for studies of compound nuclear reactions. Typical experiments include measuring differential cross sections at various angles and energies for the production of light charged-particles, including protons, alphas, tritons, deuterons, and helions, for α-, helion-, and deuteron-induced reactions. Comparison between measured yields and theoretical calculations have shed light on the reaction mechanism for charged-particle reactions and provided critical input to theoretical reaction rate calculations. Notable results include a host of level density determinations, which were obtained by comparison of experimental data to Hauser-Feshbach calculations (Voinov et al. (2007), Ramirez et al. (2015)), and assessments of the extent to which the compound and direct nuclear reaction mechanisms contribute to particular reaction cross sections (Al-Quraishi et al. (2000)).

## 5. Keck Thin Films Analysis Facility

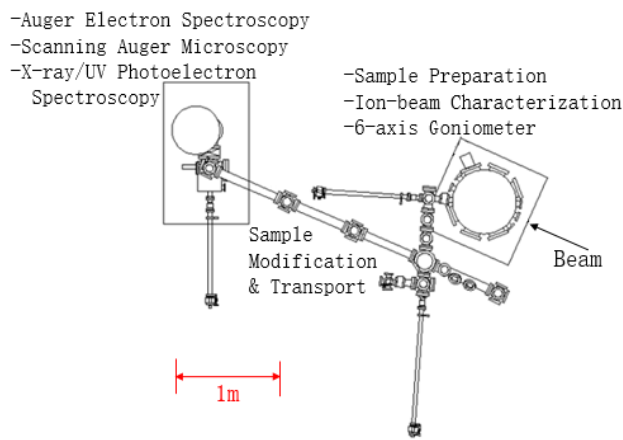

Fig. 4: W. M. Keck Thin Films Analysis Facility. (Modified from Kang (2002).)

The Keck Thin Films Analysis Facility, shown in Figure 4, is a suite of instrumentation for condensed matter physics experiments, including equipment for conducting accelerator-based material studies. The Keck Facility enables thin film creation, modification, and analysis within an ultra-high vacuum environment. Experimental equipment includes an ion beam analysis chamber, an XSAM800 device, two metalorganic chemical vapor deposition systems, and an ion transfer system (Kang (2002)). The ion beam analysis chamber is designed to enable thin films analysis via Rutherford backscattering spectroscopy, nuclear reaction analysis, and elastic recoil spectroscopy. The XSAM800 device includes equipment for X-ray and UV photoelectron spectroscopy, Auger electron spectroscopy, and scanning Auger microscopy.

The unique capabilities of the Keck Facility have been employed for a variety of studies in applied nuclear physics. The coupled target evaporation and ion beam analysis equipment allow the stoichiometry of thin films to be monitored in real time during thin-film creation, allowing, for instance, systematic studies of novel semiconductor properties (Kang and Ingram (2003), Mandru et al. (2016)). The ultra-high vacuum sample-transfer system allows thin film properties to be monitored in between systematic modifications, e.g. heating cycles, via light- and ion-based analysis techniques (Kayani et al. (2009)).

## 6. Dedicated Rutherford Backscattering Station

The Rutherford backscattering spectroscopy (RBS) station, shown in Figure 5, is a dedicated chamber for sample analysis via RBS and proton recoil detection. Two silicon surface barrier detectors, one fixed and one mounted to a goniometer, a Faraday cup, and a rotating target system are kept available for rapid sample characterization. Common applications include target thickness and composition measurements for nuclear physics studies, quantification of thin film properties, and systematic studies of ion beam damage to various materials. The RBS chamber has also been used for high-precision scattering cross section determinations to constrain optical models for nuclear reaction theory (Boukharouba et al. 1992). In addition, the RBS station is a staple for both undergraduate and graduate nuclear physics laboratory experiments.

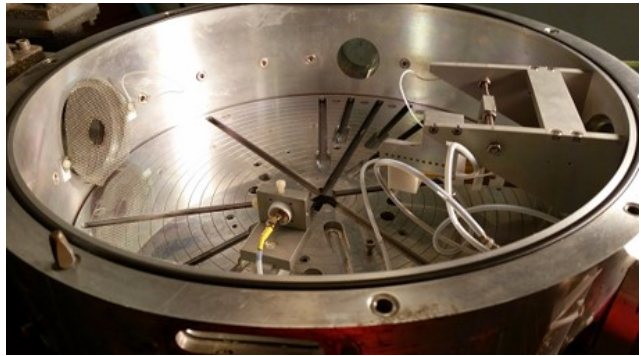

Fig.5: Dedicated Rutherford backscattering chamber.

## 7. Multipurpose Target Stations and Near-Future Upgrades

In addition to the four dedicated target stations, two target stations are presently designated for temporary experimental set-ups. These stations provide operational flexibility and obviate the need to deconstruct or alter the well-characterized existing set-ups. Several past experimental campaigns have capitalized on this flexibility to perform timely measurements on topics of current interest, including studies of the nuclear force via n-p scattering (Boukharouba et al. (2001), Boukharouba et al. (2010)) and the exploration of key resonances for nuclear astrophysics (Matei et al. (2008), Sayre et al. (2012)). Further flexibility in operations is provided by the beamline hosting the charged-particle time-of-flight chamber, as this chamber can be passed through to perform experiments further downstream. This has been done, for instance, for the commissioning of a new detector concept for homeland security applications (Micklich et al. (2003)) and for determinations of γ-strength functions for compound nuclear reactions (Voinov et al. (2010)).

The Edwards Accelerator Laboratory is continuing to develop new experimental capabilities, capitalizing on existing equipment and expertise. In order to extend neutron detection to low-yield scenarios, a detector for moderated fast-neutrons in the long-counter style is being developed. Near-future plans for this detector will be to measure (α,n) reaction cross sections in the energy-range of astrophysical interest. Additionally, an upgrade to the beam swinger target station is planned, where a silicon detector array with significant solid-angle coverage will be used to measure decays from compound nuclear states populated in two-proton transfer measurements.

**Acknowledgements**

This work was supported in part by Department of Energy Grant № DE-FG02-88ER40387 and DE-NA0002905. We also acknowledge generous support over the years from the U.S. Atomic Energy Commission, the U.S. Department of Energy, Lawrence Livermore National Laboratory, Ohio University, the National Nuclear Security Administration, the National Science Foundation, and the W.M. Keck Foundation.